\documentclass[english,preprint2,bibyear]{aa}
\usepackage[varg]{txfonts}

\usepackage{amsbsy}
\usepackage{amstext}
\usepackage{amsmath}
\usepackage{array}
\usepackage{babel}
\usepackage{color}
\usepackage[font=small,labelfont=bf]{caption}
\usepackage[T1]{fontenc}
\usepackage{graphicx}
\usepackage{natbib}
\usepackage{units}
\usepackage[varg]{txfonts}

\usepackage{natbib,twoopt}
\usepackage[breaklinks=true]{hyperref} 
\bibpunct{(}{)}{;}{a}{}{,}

\newcommand{\mbh}{M_{\bullet}}
\newcommand{\ml}{\dot{M}_{\bullet\mbox{\scriptsize{low}}}}\newcommand{\mm}{\dot{M}_{\bullet\mbox{\scriptsize{med}}}}
\newcommand{\mh}{\dot{M}_{\bullet\mbox{\scriptsize{high}}}}

\newcommand{\msun}{{\rm M}_{\odot}}

\begin{document}

\title{Unveiling Gargantua: A new search strategy for the most massive central cluster black
 holes}

\author{M.~Brockamp\inst{1}\thanks{brockamp@mpifr-bonn.mpg.de (MB), h.baumgardt@uq.edu.au (HB),
 sbritzen@mpifr-bonn.mpg.de (SB)} \and H.~Baumgardt\inst{2} 
\and S.~Britzen\inst{1} \and A.~Zensus\inst{1}}

\institute{Max Planck Institut f\"ur Radioastronomie, Auf dem H\"ugel 69, 53121 Bonn, Germany \and
School of Mathematics and Physics, University of Queensland, St. Lucia, QLD 4072, Australia}

\authorrunning{Brockamp et al.}
\titlerunning{Unveiling Gargantua}
\date{}
 
\abstract {} 
{We aim to unveil the most massive central cluster black holes in the universe.} 
{We present a new search strategy which is based on a black hole mass gain sensitive 'calorimeter' and 
which links the innermost stellar density profile of a galaxy to the adiabatic growth of its central SMBH.
In a first step we convert observationally inferred feedback powers into SMBH growth
rates by using reasonable energy conversion efficiency parameters, $\epsilon$. In the main part
of this paper we use these black hole growth rates, sorted in logarithmically increasing steps encompassing
our whole parameter space, to conduct $N$-Body computations of brightest cluster galaxies with the newly
developed \textsc{Muesli} software. For the initial setup of galaxies we use core-S\'{e}rsic models in order
to account for SMBH scouring.} {We find that adiabatically driven core re-growth is significant
at the highest accretion rates. As a result, the most massive black holes should be located in BCGs with less
pronounced cores when compared to the predictions of empirical scaling relations which are usually calibrated in
less extreme environments. For efficiency parameters $\epsilon<0.1$, BCGs in the most massive, relaxed and
X-ray luminous galaxy clusters might even develop steeply rising density cusps.
Finally, we discuss several promising candidates for follow up investigations, among them the nuclear black hole
in the Phoenix cluster. Based on our results, it might have a mass of the order of $\unit{10^{11}}{\msun}$.} {}

\maketitle
\titlerunning{Unveiling Gargantua}

%

\section{Introduction}\label{Introduction}
The detection of supermassive black holes (SMBHs) at the high-mass end of the mass
scale is of tremendous importance for constraining limitations of empirical scaling relations
and thus for the understanding of the related evolution and growth of both SMBHs and
cosmic structures like galaxies, galaxy groups or even clusters of galaxies. 

SMBH scaling relations link different properties of galaxies such as total luminosity and
bulge mass \citep{1995ARA&A..33..581K, 2002MNRAS.331..795M, 2003ApJ...589L..21M,
2004ApJ...604L..89H, 2013ApJ...764..184M}, the velocity dispersion, $\sigma$,
\citep{2000ApJ...539L...9F, 2000ApJ...539L..13G, 2009ApJ...698..198G,
2013ApJ...764..184M}, and the central surface brightness profile \citep{2007ApJ...662..808L, 
2009ApJ...691L.142K, 2013AJ....146..160R, 2014MNRAS.444.2700D} to the mass of the nuclear SMBH. Currently,
there is a debate whether the empirical $\mbh-\sigma$ relation is a consequence of black
hole feedback regulated galaxy growth (e.g. \citealt{1998A&A...331L...1S, 2012ARA&A..50..455F})
or results from stochastic averaging over numerous galaxy coalescences in a cosmological
merger tree \citep{2011ApJ...734...92J}.

The action of binary or multiple black holes after merger events of their host galaxies is
suspected to be responsible for the formation of shallow central brightness profiles of luminous
galaxies \citep{1991Natur.354..212E, 2001ApJ...563...34M, 2006ApJ...648..976M} with
V-band magnitudes $M_{V}\lesssim-21$ \citep{2007ApJ...664..226L}. This scenario is supported
by \cite{2014ApJ...782...39T} who found a tangentially biased
velocity distribution within the break radius, $R_{\mbox{\scriptsize{break}}}$, which is located at
the maximal curvature of the surface brightness profile. In the picture of SMBH merging,
$R_{\mbox{\scriptsize{break}}}$ forms out, along with a tangentially biased velocity distribution,
by the preferential scattering/removal of stars on eccentric orbits with low angular momentum.
Therefore, the tight relationship as indicated by scaling relations between $\mbh$ and properties
of the central galactic brightness profile seems reasonable.

In a few particular and extreme cases like MS0735 and Abell~85
\citep{2009ApJ...698..594M, 2014ApJ...795L..31L}, scaling relations which are
usually calibrated in local and ``ordinary'' galaxies yield widely different black hole mass
predictions by factors of $10-100$. By using the $\mbh-R_{\mbox{\scriptsize{break}}}$
\citep{2013AJ....146..160R}, $\mbh-R_{\mbox{\scriptsize{cusp}}}$ \citep{2007ApJ...662..808L}
and $\mbh-L_{\mbox{\scriptsize{def}}}$ \citep{2009ApJ...691L.142K} relations, where
$R_{\mbox{\scriptsize{cusp}}}$ is the cusp radius of a Nuker profile and
$L_{\mbox{\scriptsize{def}}}$ is the total luminosity difference between measured and
extrapolated light profile, \cite{2009ApJ...698..594M} and \cite{2014ApJ...795L..31L} found
indications for extraordinary massive SMBHs. The central black hole masses in MS0735 and
Abell~85 might be as high as $\unit{10^{11}}{\msun}$ as long as their shallow density profiles are 
carved out by black hole merging processes\footnote{At least for the BCG in Abell~85,
\cite{2015ApJ...807..136B} show that the central brightness profile can be approximated by a S\'{e}rsic
model with a small index $n$ and without a depleted core.}.
The estimates of these three scaling relations are significantly above the predictions of the
$\mbh-\sigma$ and $\mbh-L$ relations. There is tentative evidence that it is the
$\mbh-\sigma$ (and $\mbh-L$) relation which loses its predictive power (through a vertical tilt)
above $\sigma=\unit{270}{\mbox{~km~s}^{-1}}$ \citep{2013ARA&A..51..511K}.

Unrelated studies give also rise to skepticism concerning the validity of the $\mbh-\sigma$
and $\mbh-L$ relation when dealing with the most extreme central cluster galaxies: By (i)
using the fundamental plane of accreting black holes, \cite{2012MNRAS.424..224H}
found evidence for the presence of ultramassive black holes ($\mbh\geq\unit{10^{10}}{\msun}$)
being $\approx 10$ times more massive than predictions based on the $\mbh-\sigma$ and $\mbh-L$
relations. Furthermore (ii), the case for exceptionally massive black holes is strengthened by
active galactic nuclei (AGN) with quasar like powers in form of mechanical feedback but without any
detectable X-ray emission. The central active black holes in these cool core galaxy clusters must
operate at highly sub-Eddington accretion rates to avoid detection and are most likely
ultramassive \citep{2011MNRAS.413..313H}.

Central cluster galaxies which are located in the most massive (cool core) galaxy clusters are
therefore promising candidates for hosting the largest black holes in the universe.
Indeed, stellar dynamical mass measurements of even less extreme
BCGs in the nearby Coma\footnote{We note that the Coma cluster does neither
host a cool core \citep{2013ApJ...775....4S} nor is it dominated by one central
cluster galaxy. Instead, there are three dominating galaxies which are NGC~4889, NGC~4874
and NGC~4839. They will likely coalesce sometime in the future \citep{2007A&A...468..815G}.
Their merger will give rise to a black hole of the order of $\mbh \approx
\unit{5\cdot10^{10}}{\msun}$.  For this calculation we used direct SMBH mass estimates from
\cite{1998AJ....115.2285M} for NGC~4874,  \cite{2011Natur.480..215M} for
NGC~4889 and obtained a mass of $\mbh\approx\unit{6.5\cdot10^{9}}{\msun}$
for NGC~4839 with the help of the $\mbh-R_{\mbox{\scriptsize{break}}}$ relation
from \cite{2013AJ....146..160R} and a graphically evaluated break radius from
Figure~3 in \cite{2004AJ....127...24J}. Mass loss through gravitational wave
emission \citep{2008PhRvD..78h1501T} but also stellar accretion
in non-spherical potentials \citep{2013ApJ...767...18L, 2014CQGra..31x4002V} is
neglected.} and Leo galaxy cluster as well as lenticular galaxies unveiled black holes around
$\mbh=\unit{10^{10}}{\msun}$ \citep{2011Natur.480..215M, 2012Natur.491..729V}.

This paper investigates the impact of hot cluster gas accretion on the dynamics
and density profiles of BCGs. We relate the old idea of adiabatically growing black holes
\citep{1980ApJ...242.1232Y} in combination with core depletion by merging SMBHs to create a
black hole mass gain sensitive ``calorimeter''. This tool offers a way to test
SMBH accretion models, improve the understanding of scaling relations which depend on the surface
brightness profile and help to unveil the most massive cluster black holes.  If combined with
the total feedback energy of a galaxy cluster, the inner slope and break radius, $R_{\mbox{\scriptsize{break}}}$,
of its central cluster galaxy and SMBH mass, the ``calorimeter'' might be used to uniquely determine the
SMBH mass gain during the early quasar era, through subsequent merging activity and the accretion of
hot cluster gas. Furthermore, it can be used to infer the time averaged binding energy conversion
efficiency parameter, $\epsilon$.

In Section~\ref{Models} we first compile observationally motivated black hole growth
rates. These rates also cover the extremes, from a mere spin powered feedback scenario
($\mbox{d}\mbh/ \mbox{d}t\approx0$) up to SMBH growth with small $\epsilon$ in X-ray luminous
galaxy clusters ($L_{\mbox{\scriptsize{X}}}=\unit{10^{45}-10^{46}}{\mbox{~erg~s}^{-1}}$) like
the Phoenix- \citep{2012Natur.488..349M} or RX J1347.5-1145 cluster \citep{2004A&A...427L...9G}.
In the main part of this paper (Section~\ref{Computations_Results}) we perform computations with the
newly developed \textsc{Muesli} software \citep{2014MNRAS.441..150B} in order to evaluate the impact
of the (compiled) SMBH growth rates on the innermost surface brightness profiles. In Section~\ref{Discussion}
we discuss the obtained results, present a few cluster candidates which are recommended for follow-up
observations and describe future applications and improvements of the ``calorimeter''. Our main findings
are summarized in Section~\ref{Summary}.

\section{Accretion models}\label{Models}

In this section we present central cluster black hole growth models which serve as the initial setup for our
$N$-Body computations in Section~\ref{Computations_Results}. They are obtained from
the observed AGN feedback power in combination with reasonable accretion efficiency parameters
(Section~\ref{Conversion}). Four mutually independent methods are reviewed in ascending order with regard
to their mass accretion rate. Due to the vast amount of literature our compiled list is far from being complete,
but it covers the extreme scenarios. A consistency check is made in Section~\ref{A_Models}
by comparing the observationally motivated black hole growth rates with classical Bondi models
in order to obtain reasonable initial black hole masses for our $N$-Body computations.
Readers which are already familiar with deposited AGN feedback powers can directly skip 
to the new results presented in Section~\ref{Computations_Results}.

\subsection{Conversion of feedback power into the SMBH growth rate}\label{Conversion}

The deposited AGN feedback (jet) power, $P_{\mbox{\scriptsize{AGN}}}$, is converted into
cluster black hole growth rates, $\dot{\mbh}$, by using the relation:
\begin{equation}\label{fconversion}
\dot{\mbh}=\frac{\mbox{d}\mbh}{\mbox{d}t}=\left(1-\epsilon\right)\dot{M}_
{\scriptsize{\mbox{acc}}}=\frac{1-\epsilon}{\epsilon c^{2}}P_
{\mbox{\scriptsize{AGN}}}
\end{equation}
Here, $\dot{M}_{\scriptsize{\mbox{acc}}}=P_{\mbox{\scriptsize{AGN}}}\epsilon^{-1}c^{-2}$
is the accretion rate and $\epsilon$ is the binding energy conversion efficiency parameter. 
In order to account for uncertainties related to $\epsilon$, we use the six efficiency parameters,
$\epsilon=1.0/0.42/0.2/0.1/0.057/0.01$, which span over two orders of magnitude.
They are motivated as follows:
\begin{itemize}
\item $\epsilon=1.0/0.1$: These values are obtained from the
electromagnetic jet efficiency formula, $\epsilon\approx 0.002\cdot\left(1-|a| \right)^{-1}$
\citep{2006ApJ...641..103H}, in combination with black hole spin values $a=0.998$ 
and $a=0.98$. While $a=0.998$ is the theoretical limit which can be obtained in accretion
processes \citep{1974ApJ...191..507T}, $a=0.98$  corresponds to the averaged spin
of 48 luminous AGN at high redshift ($z \geq 1.5$) with masses
$\mbh\geq\unit{3\cdot 10^{9}}{\msun}$ \citep{2014ApJ...789L...9T}. Their dimensionless
spin parameter, $a=0.98$, is derived from deduced radiative efficiency parameters of
individual accretion disks. Additionally, $\epsilon=0.1$ is often used as the canonical value for
the energy conversion efficiency (e.g. \citealt{1992apa..book.....F}).
\item $\epsilon=0.42/0.2/0.057$: The value $\epsilon=0.057$ ($\epsilon=0.42$) is the maximal
conversion efficiency of a (co-rotating) circular accretion flow (within the equatorial plane) 
around a non spinning (maximally spinning) black hole (e.g. \citealt{1989pbh..book.....N}).
The parameter $\epsilon=0.2$ lies in between.
\item $\epsilon=0.01$: The kinetic jet/feedback efficiency parameter derived from AGN
observations \citep{2008MNRAS.383..277K, 2008MNRAS.388.1011M}.
\end{itemize}

\subsection{Growth rate based on observational models}\label{growth_rate_ob}

By using the empirical relationship between mechanical work and jet luminosity at 1.4 GHz
from \cite{2010ApJ...720.1066C}, \cite{2013ApJ...763...63M} obtained a time and galaxy cluster
averaged AGN power of $P_{\mbox{\scriptsize{AGN}}}=\unit{3\cdot10^{44}}{\mbox{~erg~s}^{-1}}$.
The corresponding SMBH growth rate is:
\begin{equation}\label{fmod_low}
 \ml=0.005 \left(\frac{1-\epsilon}{\epsilon}\right)\msun\mbox{yr}^{-1}
\end{equation}
Equation~(\ref{fmod_low}) defines our low mass accretion model as it is strongly exceeded by AGN in some
of the most massive cool core galaxy clusters which are the most promising candidates for hosting the
heaviest black holes.

X-ray observations of these clusters demonstrate that the deposited AGN feedback power
required to inflate X-ray cavities within the intra cluster gas (ICM) exceeds $P_{\mbox{\scriptsize{AGN}}}=
\unit{10^{45}}{\mbox{~erg~s}^{-1}}$. \cite{2012MNRAS.421.1360H} report cavity powers being as
high as $P_{\mbox{\scriptsize{AGN}}}=\unit{\left(4-8\right)\cdot10^{45}}{\mbox{~erg~s}^{-1}}$ for
individual cases of massive galaxy clusters. Due to a duty cycle (i.e. fraction of time where the central
cluster black hole is active) of at least $60\%-70\%$ \citep{2006MNRAS.373..959D, 2012MNRAS.427.3468B},
we assume a conservative value of $P_{\mbox{\scriptsize{AGN}}}=\unit{3\cdot10^{45}}{\mbox{~erg~s}^{-1}}$
which is also found in \citep{2015ApJ...805...35H}. This corresponds to
\begin{equation}\label{fmod_mid}
 \mm=0.05\left(\frac{1-\epsilon}{\epsilon}\right)\msun\mbox{yr}^{-1}
\end{equation}
which defines our medium mass accretion model. The assumption of linear mass increase is justified by
cavity energetics which seemingly do not evolve with redshift \citep{2012MNRAS.421.1360H}.

But even the value defined in Equation~(\ref{fmod_mid}) has to be regarded as a lower limit. Neither is
the feedback energy fully deposited inside the cool core nor are cavities the only source of ICM heating.
Energy can be transferred out to much larger radii via large scale shocks or sound waves. One spectacular
example is the aforementioned system MS0735 with its bipolar $\unit{200-240}{\mbox{~kpc}}$ X-ray
cavities surrounded by Mach 1.26 shocks waves \citep{2014MNRAS.442.3192V}. They extend to at
least twice the size of the cooling radius. And indeed, studies which calculate the total injected AGN
feedback energies out to large radii obtain time averaged jet powers as high as $P_{\mbox{\scriptsize{AGN}}}=
\unit{10^{46}}{\mbox{~erg~s}^{-1}}$ \citep{2011ApJ...738..155M, 2012ApJ...759...87C}. These
values are deduced from comparing observed potential energies of virialized massive clusters
($M_{\mbox{\scriptsize{vir}}}\approx\unit{10^{15}}{\msun}$) to those obtained from simulations without feedback.
Following \cite{2011ApJ...738..155M}, the energetic demands and high accretion rates of these central cluster
black holes can not be lowered by splitting them into multiple sources. In most cases, X-ray cavities and shocks
are associated with the central BCG and not its satellites. Significant pre-heating of baryons prior to galaxy cluster
assembly is also disregarded by \cite{2011ApJ...738..155M}.

Following \cite{2002MNRAS.332..729C}, similar (or even higher) AGN feedback powers are
also required for balancing the self-cooling of the gas in the most X-ray luminous cool-core clusters approaching
$L_{\mbox{\scriptsize{X}}}\approx\unit{10^{46}}{\mbox{~erg~s}^{-1}}$  like MACSJ1447.4+0827
\citep{2012MNRAS.421.1360H}, RX J1347.5-1145 \citep{2004A&A...427L...9G} or the Phoenix cluster
\citep{2012Natur.488..349M}. Integrated over galaxy cluster lifetimes of seven to eight billion years and by
assuming various efficiency parameters (from as high as unity for black hole spin powered models down to a few
percent for Schwarzschild black holes, see Section~\ref{Conversion}) this results in up to
$\Delta\mbh\approx\unit{10^{10}}{\msun}$ of accreted mass. Feedback energies as high as 
$P_{\mbox{\scriptsize{AGN}}}=\unit{10^{46}}{\mbox{~erg~s}^{-1}}$ define our high mass accretion scenario:
\begin{equation}\label{fmod_high}
 \mh=0.18\left(\frac{1-\epsilon}{\epsilon}\right)\msun\mbox{yr}^{-1}
\end{equation}

\begin{figure}[t!!]
  \centering
  \makebox[-2cm]{\includegraphics[width=0.52\textwidth]{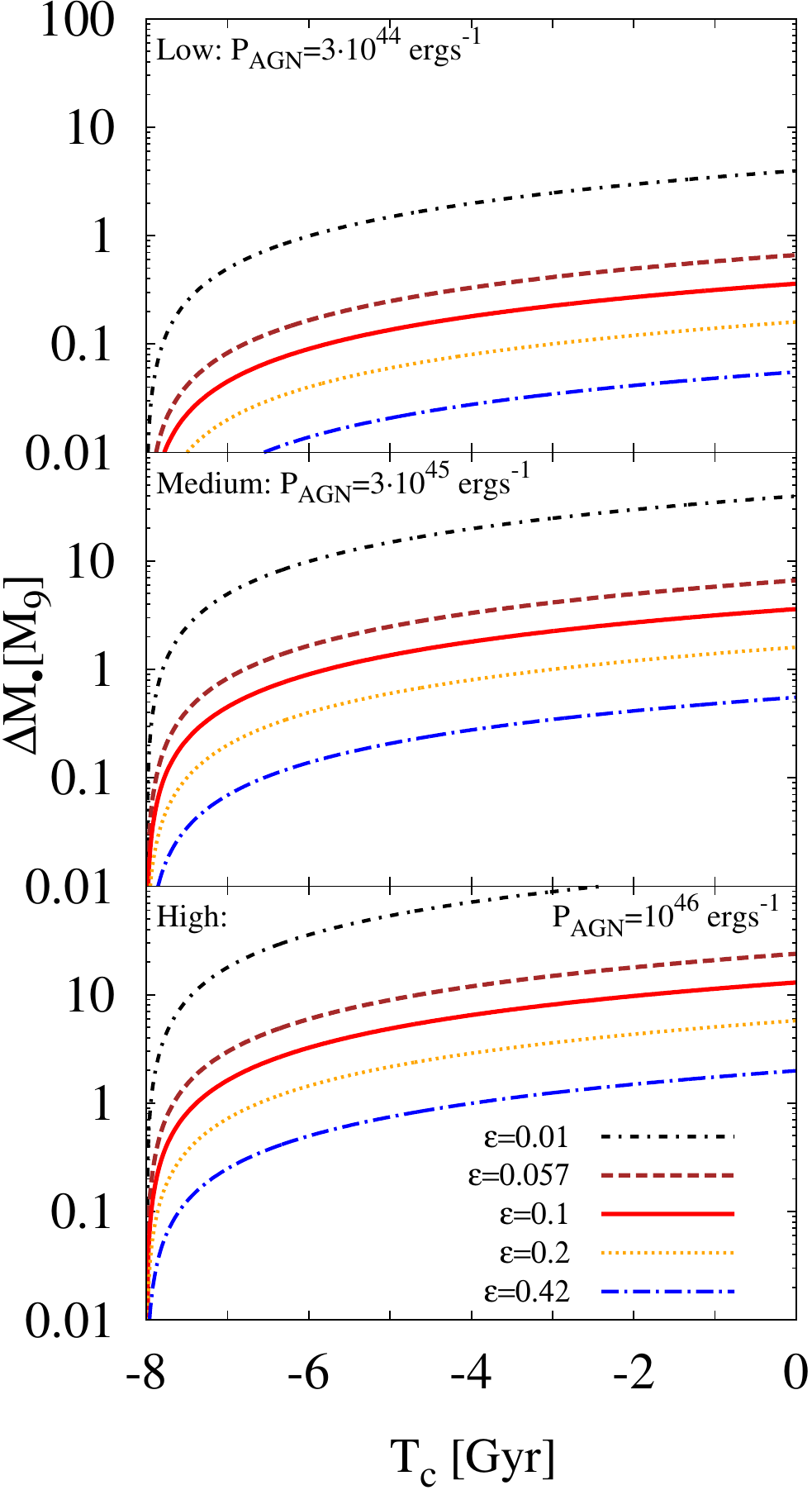}}
  \caption{The SMBH growth rate is obtained from the central 
cluster AGN feedback power which is constrained by means of
different strategies and cluster samples. The colors and line
types indicate the binding energy conversion efficiency
parameters which must be adopted for the conversion
(Equation~\ref{fconversion}). The cosmic time is defined to
be $T_{c}\left(z=0\right)=0$ and we assume that feedback/accretion
started operating at redshift $z=1$ (8 billion years ago). SMBH
masses can be obtained from the relation $\mbh\left(t\right)=
M_{\bullet \mbox{\scriptsize{init}}}+\Delta \mbh\left(t\right)$.}
\label{Fig_Growth_Models_ob}
\end{figure}

Central cluster black holes growing at such prodigious rates might pose a challenge for theories
which assume that significant SMBH growth ceased early on in accordance with cosmic downsizing
of the AGN activity \citep{2005A&A...441..417H, 2013ApJ...764...45K}. Alternatively, the 
feedback might be powered mainly by trapped spin energy ($\epsilon\approx1$) or central cluster
black holes might have reached dimensions where even $\Delta\mbh\gtrsim\unit{10^{9}}{\msun}$
of accreted gas represents only a tiny fraction of their total mass \citep{2012MNRAS.421.1360H}.
We note that the latter idea is in accordance with the existence of individual cases of extremely
massive AGN in the early universe \citep{2009MNRAS.399L..24G, 2014ApJ...789L...9T, 
2015MNRAS.447.3368B, 2015Natur.518..512W, 2015ApJ...806..109J} and is conform with the
upper limit of the quasar/black hole mass function (BHMF) (\citealt{2013ApJ...764...45K}, their Chapter~3.3).
The obtained BHMF is calibrated by the virial mass estimates of $~58.000$ Type 1 quasars. 

The observation-based models are graphically illustrated in Figure~\ref{Fig_Growth_Models_ob}.
Note that models with $\epsilon=1$ ($\Delta\mbh=0$) are not plotted subject to the
logarithmic spacing of the y-axis. With regard to the enormous spread in
$\epsilon$ (Section~\ref{Conversion}), the accumulated SMBH mass over 8 Gyr (i.e. the typical
lifetime of galaxy clusters), varies from $\Delta\mbh=0$ up to several $\unit{10^{10}}{\msun}$.
Consequently, the growth of the SMBH can not be uniquely constrained from the deposited feedback energy.
Therefore, it is important to understand the influence of the growing black hole on its immediate
surroundings. This leads us to the idea of the black hole mass gain sensitive ``calorimeter" which is part of
Section~\ref{Computations_Results}.

\subsection{Consistency check: Classical Bondi models}\label{A_Models}
Through a comparison with theoretically justified models we can explore whether
the initial black hole masses which are chosen for our $N$-Body computations and which can
not be directly deduced from the observed feedback powers are realistic.
We focus on galaxy clusters with relaxed gas atmospheres and ignore cold gas accretion
scenarios subject to the absence of a positive correlation between jet power and molecular
gas mass \citep{2011ApJ...727...39M}. We assume that accretion flows are not disturbed
through shock heating by collisions and consider only well established models, i.e. classical Bondi flows
\citep{1952MNRAS.112..195B} with some improvements from more recent investigations.
The usage of sophisticated (analytical) models like radiating Bondi flows \citep{2012ApJ...754..154M}
is left for future studies. 

\begin{figure}[t!]
  \centering
  \makebox[-2cm]{\includegraphics[width=0.505\textwidth]{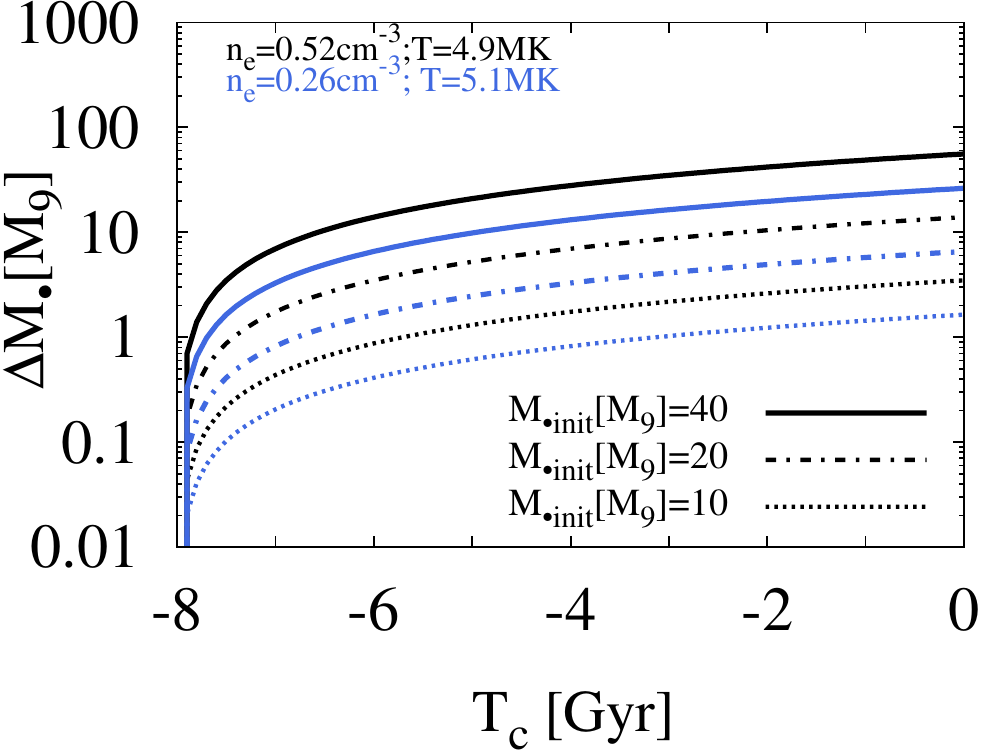}}
  \caption{Black hole growth models related to the classical Bondi accretion scenario. Although,
the rates differ widely and predict extreme upper values, they lie within the parameter range
of the observation-based models which are plotted in Figure~\ref{Fig_Growth_Models_ob}. In
case of very massive seed black holes around $\mbh=\unit{10^{10}}{\msun}$ at $z=1$,
Bondi accretion alone is sufficient to explain the deposited feedback energies even in 
the most extreme cool core clusters.}
\label{Fig_Growth_Models_bondi}
\end{figure}

SMBH growth rates related to classical Bondi accretion models are shown in Figure~\ref{Fig_Growth_Models_bondi}.
We do not consider the influence of the growing black hole on the accretion rate but
discuss some consequences this might have in Section~\ref{CAO}. Our setup includes three initial
black hole masses with $\mbh=1,2,4\cdot\unit{10^{10}}{\msun}$. The highest value corresponds
to the upper quasar mass obtained from the black hole mass function \citep{2013ApJ...764...45K}.
Individual cases of black holes with such masses at high redshift are also found in \cite{2014ApJ...789L...9T}
and \cite{2015ApJ...806..109J} and are required to successfully model (but not strictly unambiguously)
the spectral energy distribution and total luminosity of the blazar S5 0014+813 at z = 3.37
\citep{2009MNRAS.399L..24G}\footnote{Note that blazars are highly beamed sources, so for every
blazar there should be several non-beamed AGN with similar properties \citep{2009MNRAS.399L..24G}.}.
The three initial SMBH masses are discriminated by using different line-types in Figure~\ref{Fig_Growth_Models_bondi}.

Apart from some modifications which are specified in the text, we make use of the same strategy
as \cite{2006MNRAS.372...21A} and use their equations and parameter setting for calculating the black
hole growth rate through Bondi accretion:
\begin{equation}\label{formula_bondi_classic}
 \dot{M}_{\bullet B}=\eta\left(1-\epsilon\right)4\pi\lambda\left(G\mbh \right)^{2}c_{s}^{-3}\rho
\end{equation}
Here, $\eta=0.2$ is a conservative correction parameter for rotating and viscous gas
atmospheres \citep{2011MNRAS.415.3721N}. The term $\left(1-\epsilon\right)=0.9$ guarantees
energy conservation (assuming $\epsilon=0.1$) and the coefficient $\lambda=0.25$ is related to the 
adiabatic index $\gamma=5/3$. The sound velocity at the characteristic Bondi or accretion
radius, $R_{a}=2G\mbh/c_{s}^{2}$, is $c_{s}=\sqrt{\gamma k_{B}T/ \left(\mu m_{p} \right)}$.
The parameter $ k_{B}$ is the Boltzmann constant, $T$ is the gas temperature, $\mu=0.62$ is the average
atomic weight and $m_{p}$ is the mass of the proton. The gas density, $\rho=1.13n_{e}m_{p}$, 
which is also taken from \cite{2006MNRAS.372...21A} can be calculated from the measurable
electron density, $n_{e}$.

The remaining free parameters, $n_{e}$ and $T$, which are required for our growth models are taken
from a cluster sample average. To be more precise, we calculate the arithmetic mean values of
$n_{e}$ and $T$ at $R_{a}$ from Table~4 in \cite{2013MNRAS.432..530R}. Their sample includes
the same galaxy clusters as \cite{2006MNRAS.372...21A} plus four additional objects.
Two sets of parameters (distinguished by different colors in Figure~\ref{Fig_Growth_Models_bondi}) are chosen.
The first one, $n_{e}=\unit{0.52}{\mbox{cm}^{-3}}, T=\unit{4.9\cdot10^{6}}{\mbox{K}}$, represents the mean, whereas the second
one, $n_{e}=\unit{0.26}{\mbox{cm}^{-3}}, T=\unit{5.1\cdot10^{6}}{\mbox{K}}$ is obtained from the upper ($T+\Delta T$) and lower
($n_{e}-\Delta n_{e}$) $1\sigma$ mean values. In this way, we tend to underestimate the true SMBH growth rate. 

Uncertainties related to $R_{a}$ (which depends on the assumed $\mbh$) and the parameters
$n_{e}\left(R_{a}\right)$ and $T\left(R_{a}\right)$ are compensated by a constant ratio
$c_{s}^{-3}\rho\propto PT^{-5/2}=$ const (in Equation~\ref{formula_bondi_classic}) which is
independent of radius, $r$, for $\gamma=5/3$, $P=nk_{B}T$ and $T^{\gamma}P^{1-\gamma}=$ const.
It is only important that $n_{e}$ and $T$ are measured deep within the potential of the central host galaxy because
of its influence on the Bondi accretion flow. 

Evidently, all analytical-based models which are presented in Figure~\ref{Fig_Growth_Models_bondi} lie within the
parameter range of the observation-based models (Section~\ref{growth_rate_ob}). This indicates that Bondi accretion
alone is sufficient for powering the most extreme central cluster AGN hosting the largest SMBHs. 
Furthermore, it shows that initial black hole masses as high as $\mbh=\left(1-4\right)\cdot\unit{10^{10}}{\msun}$
at the onset of hot cluster gas accretion are reasonable.

\section{Cusp formation due to SMBH growth}\label{Computations_Results}

Initial conditions and results of $N$-Body computations are presented in this section. 
The idea is to numerically investigate, for a sample of representative black hole growth
models (Section~\ref{Models}), the adiabatic response of the red/old stellar density profile
of the host galaxy to the mass which becomes trapped by the black hole. Under the restriction
that the duration of the SMBH growth phase is larger than the orbital timescale, the term 'adiabatic'
refers to the increase of the central density while simultaneously the angular
momentum distribution remains conserved. This idea goes back to \cite{1980ApJ...242.1232Y}
and was later used by \cite{1999AJ....117..744V} to explain and predict the observed power law cusps of early
type galaxies and central SMBH masses. \cite{1999AJ....117..744V} assumed
that initial galaxies already started with constant density cores (similar to globular clusters without core-collapse)
and concluded that less-luminous galaxies with steeply rising slopes are those galaxies which host the most massive
black holes when normalized to the mass of the prior constant density core. This leads to wrong
predictions (e.g. by a factor 30 for NGC~4889) because it underestimates the merger history of these
galaxies. Instead, SMBH scouring leads in a natural way to the depletion of galactic centers \citep{1991Natur.354..212E,
2001ApJ...563...34M, 2006ApJ...648..976M} and it predicts that the most massive
black holes reside in galaxies with the most pronounced cores. \cite{2002ApJ...566..801R} treated both
scenarios (adiabatic and SMBH scouring) as competing models and argued that black hole merging is superior
in explaining the observed parameter relations. Today, adiabatic black hole growth models have disappeared
mostly from the literature\footnote{During our referee process \cite{2015arXiv150806409J}
presented new adiabatic growth models by using classical S\'{e}rsic bulges for the initial setup of galaxies. Their
models cover the other extreme -- the least luminous galaxies with nuclear star clusters.}.

Motivated by the wide distribution of the central cluster black hole growth rates (Section~\ref{Models}), we
hope to rehabilitate adiabatic SMBH growth models - but this time not as a competitor model to SMBH scouring.
The relocation of mass i.e. adiabatic black hole growth \citep{1980ApJ...242.1232Y, 1984MNRAS.207..511G, 1999AJ....117..744V}
leads to a cusp regeneration process of the depleted density centers after SMBH scouring.
Of course, the obtained results depend on initial conditions and the accretion rate of hot cluster gas.
In this sense, our findings are not free of systematics, too, as we also have to make assumptions about initial conditions which
might turn out to be inappropriate. However, initial conditions can be modified in future studies. The method itself allows 
to construct a black hole mass gain sensitive ``calorimeter" which is based on the innermost surface
brightness profile and which can be used to estimate (independently of $\epsilon$) the amount of swallowed mass in 
galaxy clusters.

As already motivated in Section~\ref{A_Models}, we must assume that black hole growth (through hot gas)
is powered by accretion flows which extend far outward, e.g. classical Bondi or giant advection dominated
flows. Accretion from a nearby molecular gas reservoir which is not replenished can not lead to adiabatic cusp
regrowth owing to an unchanged potential. Star formation which can also increase the central surface brightness
is neglected and therefore our results only apply to the density profile of old stars.
Furthermore, we use a simplified merger history and assume that core profiles are created by the currently most
accepted model, the SMBH binary evolution scenario. In the scenario of cosmic structure formation, several major
merger events (4-5) are expected to occur between $z=0$ and $z=3$ \citep{2007IAUS..235..381C} with a
strong decline at $z=1$. Our models start at this redshift and we assume that core-profiles were fully grown
up to $z=1$, so we can use current parameter relations for fixing the break radius. We neglect the $\approx1$
major merger events since $z=1$. Alternatively, we assume that alternating phases of core creation through
merging and adiabatic cusp regrowth are dynamically equivalent.

\subsection{$N$-Body Setup}\label{Setup}
\subsubsection{Runs}
Nine representative SMBH growth models from Section~\ref{Models} are selected for our $N$-Body
computations. On a logarithmic scale, they equally cover the whole parameter space including the
extreme scenarios. The models are listed with their initial parameters in Table~\ref{Model_parameters}.
The most extreme growth mode (Model~9) is computed by means of $N$-Body computations only from redshift
$z=1$ up to $z=0.1$. In this way, we enlarge the number of promising clusters hosting such massive SMBHs
within our light-cone.

Fully spin powered models with $\epsilon=1$ (i.e. $\Delta\mbh=0$) are automatically computed in form of 
our reference models. They do not lead to adiabatic cusp regrowth since all binding energy is radiated away.  
Note that the conversion of given cosmic epochs into redshifts is done with the public available ``Cosmological
Calculator for the Flat Universe" from Nick Gnedin by using $\Omega_{0}=0.3036$ and $H_{0}=
\unit{68.14}{\mbox{km~s}^{-1}\mbox{Mpc}^{-1}}$.

\begin{table*}[t!]
\begin{center}
\caption{Initial parameters of the model galaxies} 
\label{Model_parameters}
\begin{tabular}{ c c c c c c c c }  
\hline\hline  
Model & T[Gyr]& init.$\mbh$[$M_{9}$] & $M_{\scriptsize{\mbox{dyn}}}$[$M_{9}$] & $R_{e}$[kpc] &  $R_{b}$[kpc] &  $\Delta\mbh$[$M_{9}$] & Comment  \\ 
\hline
1 & 8 & 1.5  & 400 &  4 & 0.15  & 0.06 & OB-low, $\epsilon=0.42$ \\
2 & 8 & 5  & 770 & 7  & 0.4 & 0.17 & OB-low, $\epsilon=0.2$ \\
3 & 8 & 10  & 1550 & 18  & 0.8 & 0.38 & OB-low, $\epsilon=0.1$ \\
4 & 8  &10 & 1550 & 18  & 0.8 & 0.7 & OB-low, $\epsilon=0.057$ \\
5 & 8 &15 & 1750& 25  & 1.3 & 4 & OB-medium, $\epsilon=0.1$  \\
6 & 8  &20 & 1750& 25  & 1.8 & 6 &   OB-high, $\epsilon=0.2$ \\
7 & 8 &20  &1750 & 25 & 1.8 & 13 &  OB-high, $\epsilon=0.1$\\
8 & 8&40  &2000 & 68  & 3.8 & 40 &  OB-medium, $\epsilon=0.01$  \\
9 & 6.7 &40  & 2000 & 68 & 3.8 & 120 & OB-high, $\epsilon=0.01$ \\
\hline
\end{tabular}
\tablefoot{List of parameters selected for our models/$N$-Body computations. ``OB" stands for
observation-based models (Section~\ref{growth_rate_ob}). Slope values 
$\gamma=0.2$, $\alpha=4.25$ and $n=4$ (Equation~\ref{f_core_sersic}) are used for all models.
The huge initial SMBH masses in excess of $\unit{10^{10}}{\msun}$ are motivated at the end of
Section~\ref{growth_rate_ob} and in Section~\ref{A_Models}.
While dynamical masses, $M_{\scriptsize{\mbox{dyn}}}$, and effective radii, $R_{e}$, of our model
galaxies are constrained by representative galaxies, the break-radius, $R_{b}$, is taken from the
scaling relation in \cite{2013AJ....146..160R}. SMBH masses during the $N$-Body computations are
linearly increased up to $\Delta\mbh$ over the typical galaxy cluster lifetime of $T\approx 8$ Gyr. 
The $\Delta\mbh$ values are selected to equally cover (in log-space) the whole parameter range of SMBH
growth rates which are specified in Section~\ref{Models}. Model~9 is computed from $z=1$ until $z=0.1$.
In this way the amount of potential candidates hosting extraordinary massive SMBHs is increased over a sufficiently
large light-cone.} 
\end{center}
\end{table*} 

\subsubsection{Initial galaxy profiles and parameters}
We use core- S\'{e}rsic models \citep{2003AJ....125.2951G, 2004AJ....127.1917T} for our initial galaxy
setup since we assume that galactic cores are created within the first 5-6 Gyr, prior to the onset of AGN 
cluster feedback. The models are generated in equilibrium (including central black holes)
using the procedure described in \cite{2014MNRAS.441..150B}. In order to exclude any dynamical instabilities
related to anisotropic models (i.e. tangetially biased within $R_{\mbox{\scriptsize{break}}}$ and radially
biased at large radii), we only use models with an isotropic velocity distribution. Following
\cite{2003AJ....125.2951G} and \cite{2004AJ....127.1917T}, the six parametric Core- S\'{e}rsic models have a light
profile, $I\left( r \right)$, (i.e. 2D density profile assuming a constant mass to light ratio) of the form:
\begin{align}\label{f_core_sersic}
 I\left( R \right) =& I_{b}2^{-\frac{\gamma}{\alpha}}  \exp\left[   b\left(  2^{\frac{1}{\alpha}}R_{b}
R_{e}^{-1} \right)^{\frac{1}{n}}     \right] \times  \\
& \left[1+ \left(\frac{R_{b}}{R}\right)^{\alpha} \right]^{\frac{\gamma}{\alpha}}\exp\left\{
-b\left[  \frac{R^{\alpha}+ R_{b}^{\alpha}}{R_{e}^{\alpha}}  \right]^{\frac{1}{n\alpha} }\right\}.\nonumber
\end{align}
Here, $b$ is a normalization parameter such that half of the total luminosity is produced within the
projected half-light radius, $R_{e}$. The parameter, $b$, is obtained by numerically solving the integral equation~A10
in \cite{2004AJ....127.1917T}. The S\'{e}rsic index, $n$, specifies the light concentration of the outer profile
beyond the break radius, $R_{\mbox{\scriptsize{break}}}$, (for short $R_{b}$) where the profile transforms
into a power-law with slope $\gamma$. The parameter $\alpha$ controls the sharpness of the transition and $I_{b}$
is the light intensity at $R_{b}$. Selected parameters for our models are given in Table~\ref{Model_parameters}
and are discussed below:

By first fixing the initial SMBH mass to be $\mbh[M_{9}]=1.5,5,10,15,20,40$, $R_{b}$ is then extracted from
the parameter relation in \cite{2013AJ....146..160R}. Although values as high as $R_{b}=\unit{3.8}{\mbox{ kpc}}$
seem far-fetched, there exist central cluster galaxies e.g. Abell~2261 and Abell~85 which have similar or even larger 
break-radii \citep{2012ApJ...756..159P, 2014ApJ...795L..31L}. The effective radii, $R_{e}$, and dynamical
masses, $M_{\scriptsize{\mbox{dyn}}}$, of our model galaxies are based on real galaxies (NGC~1399, M~60,
NGC~3842, NGC~4889, NGC~6166) with similar black holes \citep{1998AJ....115.2285M, 2007ApJ...664..226L,
2013ApJ...764..184M}. The SMBH masses ($\mbh[M_{9}]=15,20$) are within the one sigma confidence limits of
NGC~4889. Hence, NGC~4889 is used as a representative for our highest mass models. For the uppermost value
($\mbh[M_{9}]=40$) we use NGC~6166. So far, it is the only source with a measured black hole mass based on
stellar kinematics which is comparable to that limit \citep{1998AJ....115.2285M}. However, its mass has not yet
been confirmed by other studies. For galaxy models 8 $\&$ 9 the bulge mass of NGC~6166 is rounded upwards to
$M_{\scriptsize{\mbox{dyn}}}[M_{9}]=2000$.

We use the central slope parameter $\gamma=0.2$ which is assumed to be shaped by SMBH merging. This value is in
accordance with several observed core-type galaxies in \cite{2013AJ....146..160R} and \cite{2014MNRAS.444.2700D} 
and it results in a 3D density profile which declines faster than $r^{-0.5}$. This is an important constraint for isotropic models
as it guarantees the dynamical stability within the influence radius of the black hole \citep{1994AJ....107..634T}.
The transition parameter, $\alpha=4.25$, is obtained from the arithmetic mean values from all galaxies in
\cite{2013AJ....146..160R} and \cite{2014MNRAS.444.2700D} except NGC~7768. This particular galaxy has a
too sharp transition value, $\alpha \rightarrow\infty$, which would otherwise bias the arithmetic mean. All models and their
parameters are summarized in Table~\ref{Model_parameters}.

\begin{figure*}[t!!]
  \centering
  \makebox[2cm]{\includegraphics[width=1.0\textwidth]{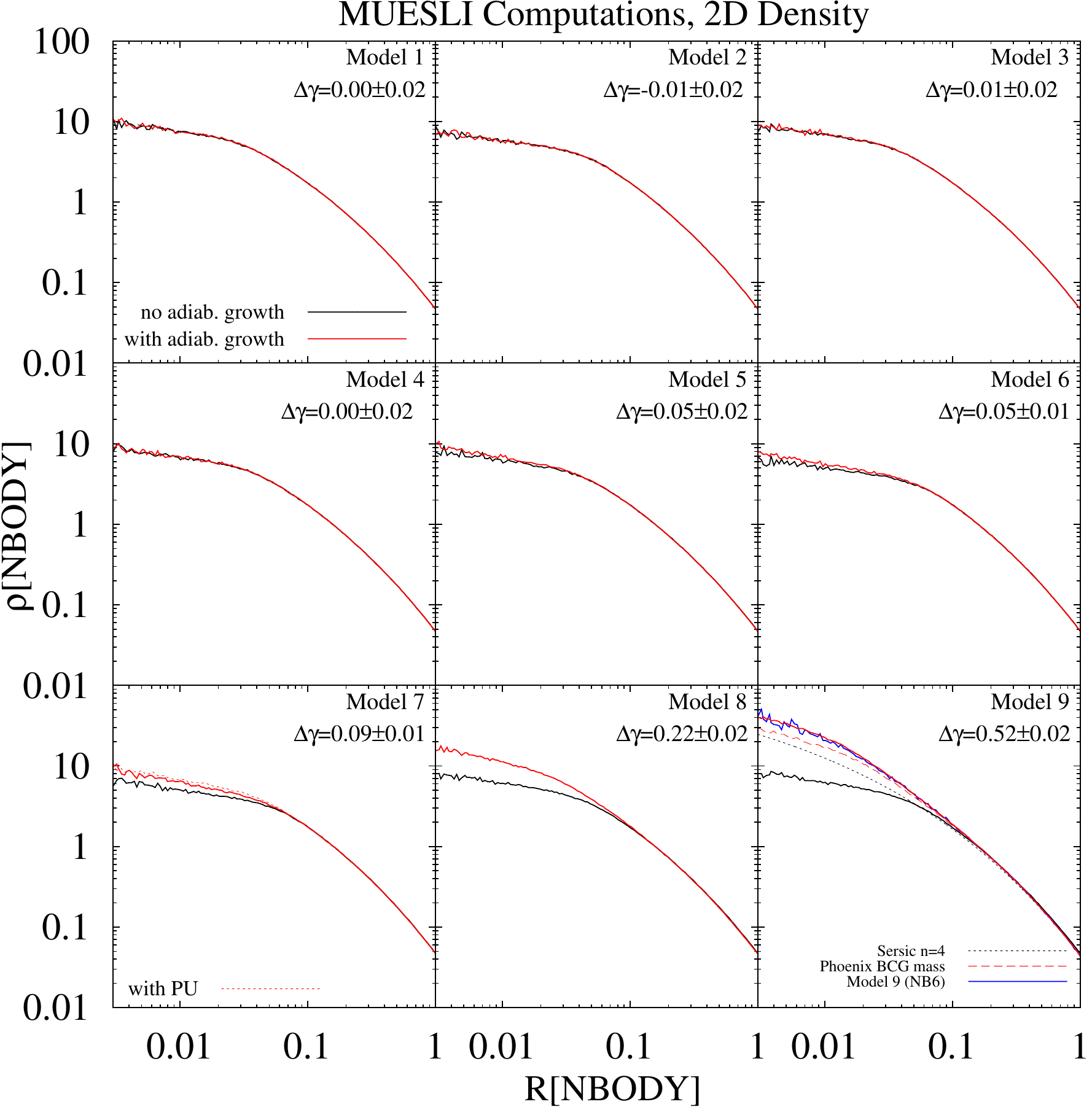}}
  \caption{Projected (2D) density profiles of all nine models.
 Red lines represent adiabatic SMBH growth models after $T=8$~Gyr (Models~1-8) and $T=6.7$~Gyr (Model~9)
 whereas black lines show reference models without growing SMBHs (i.e. $\Delta\mbh=0$). The density slopes, $\gamma$, are fitted 
within half the size of the initial break radii by using a simple power law approximation. Hence, they are influenced by the outer profile and
we quote them with their sign. In ascending order of the computed reference Models~1-9 these are:
$\gamma_{\mbox{\scriptsize{ref}}}=-0.26;-0.26;-0.3;-0.3;-0.28;-0.23;-0.25;-0.3;-0.3$.
The quantity $\Delta\gamma=\gamma_{\mbox{\scriptsize{ref}}}-\gamma_{\mbox{\scriptsize{adiab}}}
 \geq 0$ results from a growing density cusp due to adiabatic SMBH growth.
 Evidently, accretion rates with $\Delta\mbh \gtrsim \mbh\left( t_{0} \right)$  lead to a cusp regeneration process.
 Smaller accretion rates (Models~1-5) have no strong impact on the innermost 2D density slope.
 In panel~9 a pure Sersic $n=4$ model is added for comparison (black dotted line). The solid-blue line corresponds
to the \textsc{NBODY6} computation of the same model. It is in excellent agreement with the 
solid-red line. The red-dashed line is similar but is scaled to a 1.5 times larger host galaxy mass, similar to the
Phoenix BCG \citep{2013ApJ...765L..37M}.}  
\label{Dens_muesli}
\end{figure*}

\subsubsection{$N$-Body computations}
The main computations are performed with the \textsc{Muesli} software for collision-less dynamics \citep{2014MNRAS.441..150B}
on six nodes of the VLBI-computer cluster at the Max Planck Institute for Radio Astronomy. The calculations lasted more than a month. 
\textsc{Muesli} is a multi-purpose platform for the simulation of elliptical galaxies, their nuclear black holes and (not relevant for
this paper) globular clusters systems. As a check we also performed a few computations by direct N-body simulations using a
self-customized version of \textsc{NBODY6} with $N=5\cdot10^{5}$ particles \citep{1999PASP..111.1333A,
2003gnbs.book.....A, 2012MNRAS.424..545N}. This dual strategy allows to exclude potential error sources. These are either
related to two-body relaxation processes (relevant for direct summation codes) where too small particle numbers would lead
to a cusp regeneration process around a central black hole \citep{1976ApJ...209..214B} or potential fluctuations induced by
discreteness noise in the self-consistent field (SCF) method \citep{1992ApJ...386..375H}. The SCF formalism is implemented into
the \textsc{Muesli} software. We carefully tested (by finding a converging solution) the required particle number, N, and the required
order of the radial and angular base functions. We use $10^{7}$ particles for the \textsc{Muesli} simulations. Furthermore, in order
to decrease the computational cost and to reduce numerical errors (i) Model~1 (\&~2) are not evolved for 850 (520) i.e. 8 Gyr but only
for 50 $N$-Body time units ($\Delta\mbh$ remains unaffected). This strategy is applicable as we are interested in adiabatic processes
only and want to suppress relaxation. (ii) During computations with \textsc{Muesli}, the overall potential/density is not updated in
the Models~1-6  in order to reduce relaxation effects caused by discreteness noise.
Model~7 is computed with (red-dashed line) and without (red-solid line) potential updates (PU). It is the model where back-reaction 
effects of the changing density profile (due to SMBH mass growth) start to become important. Therefore,  Models~8 \&~9 where
$\Delta\mbh$ is of the order of the initial black hole mass itself are only computed with frequent updates of the potential.\\ 

Despite dynamical processes like SMBH wandering, dynamical heating and relaxation driven cusp formation which occurs in 
\textsc{NBODY6} computations (see e.g. \citealt{2011MNRAS.418.1308B}), the agreement with the \textsc{Muesli}
simulations is nevertheless very good. The outcome of a \textsc{NBODY6} computation is included in Figure~\ref{Dens_muesli}
(Model~9, solid-blue line) for a comparison. To exclude (as much as possible) even unknown systematics we
also compute all models without adiabatic black hole growth (i.e. with $\Delta\mbh=0$) and use them as a reference and for
comparison. They trivially correspond to fully spin powered models, too. Finally, the amount of swallowed particles by the
adiabatically growing black hole ranged from insignificant to small values when compared with the gas accretion rates and is
therefore not discussed.

\subsection{Results}\label{Results}
Figure~\ref{Dens_muesli} shows the dynamical response of the projected (2D) density profiles to adiabatically growing SMBHs obtained
with the MUESLI software \citep{2014MNRAS.441..150B}. We use dimensionless model 
units, $R, \rho_{2D}$, for a better comparison. These models can be scaled to physical units by $R_{\mbox{\scriptsize{phys}}}=1.34R_{e}R$ and
$\rho_{2D}\left( R_{\mbox{\scriptsize{phys}}} \right)=\rho_{2D}\left( R \right)M_{\scriptsize{\mbox{dyn}}}/
\left(1.34 R_{e}\right)^{2}\propto\mu\left( R_{\mbox{\scriptsize{phys}}}\right)$, where $\mu$ is the surface brightness
profile using an appropriate mass to light ratio. All computations (red curves for adiabatic models and black curves for
reference models without growing SMBH) contain $10^{7}$ particles. The resolution scale in Figure~\ref{Dens_muesli} is
$R_{\mbox{\scriptsize{phys}}}=\unit{16}{\mbox{~pc}}$ (Model~1) to $R_{\mbox{\scriptsize{phys}}}=\unit{270}{\mbox{~pc}}$ (Model~9).
The central density slope, $\gamma$, is measured within half the initial break radius. Obtained uncertainties
are based on assumed Poisson errors. Significant cusp regrowth is defined over the quantity 
$\Delta\gamma=\gamma_{\mbox{\scriptsize{ref}}}-\gamma_{\mbox{\scriptsize{adiab}}}$ which must be larger
than the sum of the individual slope errors. Evidently, noticeable cusp growth is only observed in Models~5-9
which means that the black hole accreted more than 25\% of its initial mass. 

\begin{figure}[t!!]
  \centering
  \makebox[-2cm]{\includegraphics[width=0.505\textwidth]{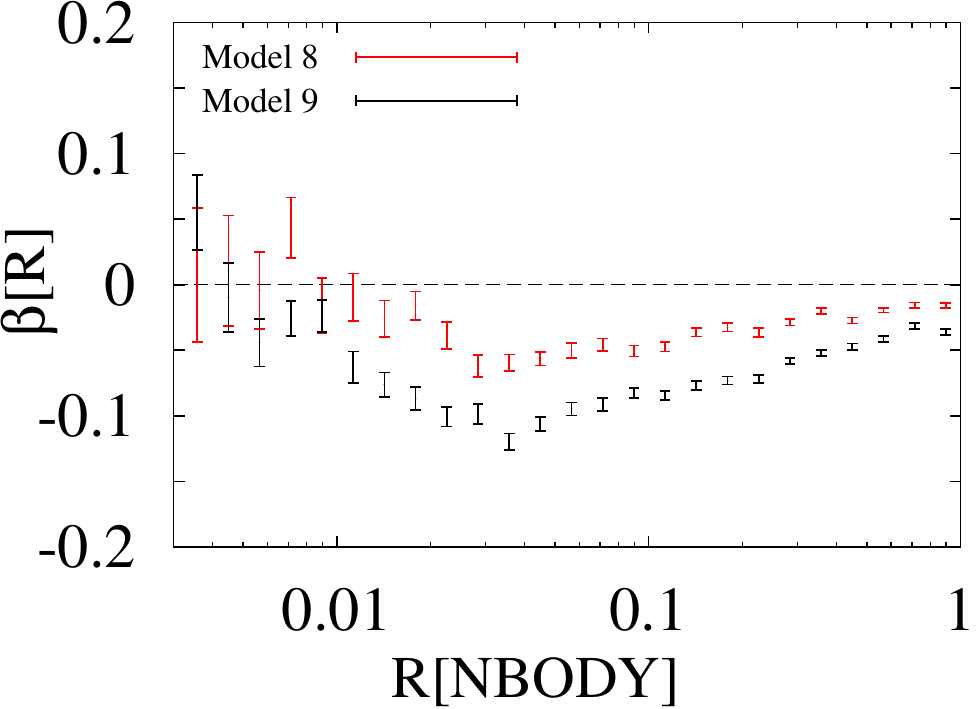}}
  \caption{The (analytical) initial (i.e. $\Delta\mbh[M_{9}]=0$, black-dashed line) and final velocity
   distribution, $\beta=1-(\sigma_{\theta}^{2}+\sigma_{\phi}^{2})/2\sigma_{r}^{2}$, of our most
   extreme Models~8 \& 9. We use the 3D radius for the evaluation of the velocity dispersion
    components, $\sigma_{\theta}, \sigma_{\phi}, \sigma_{r}$. Error values are obtained from the bootstrapping method.
    Particles on eccentric orbits are attracted more strongly towards the innermost center such that circular orbits, i.e. a
    tangetial biased region develops slightly outside but still very close to the center. This effect is relevant for the stellar
    dynamical mass measurement of the central black hole.}
\label{Fig_beta}
\end{figure}

In Figure~\ref{Fig_beta} we plot the velocity distribution, $\beta=1-(\sigma_{\theta}^{2}+\sigma_{\phi}^{2})/2\sigma_{r}^{2}$,
of Model~8 \& 9. In very good agreement with the outcomes of semi-analytical approaches
(e.g. \citealt{1984MNRAS.207..511G}, \citealt{1995ApJ...440..554Q}, their figure~2) the isotropic velocity distribution around
the pre-existing seed black hole, $\beta\left( R \right)=0$, changes. Particles on eccentric orbits are dragged more
efficiently towards the innermost center where the velocity distribution stays isotropic. Slightly outside, a tangential biased region
which is dominated by particles on more circular orbits forms. This has consequences for the stellar dynamical black hole mass
determination method. Extreme cool-core clusters like the Phoenix- or RX J1347.5-1145 cluster with rapidly growing SMBH masses
(unless $\epsilon>0.1$) are rare within our light-cone. At huge cosmic distances, the innermost isotropic region
(if it has not already been destroyed by SMBH merging) might be out of the resolution scale and what is left
is the region where circular orbits dominate. The measurable radial velocities will be biased towards lower values and thus give rise to
an underestimation of the true black hole mass if not properly taken into account. The effect is enhanced, if the initial model already
starts with a tangentially biased core as expected from SMBH scouring models.
\section{Discussion}\label{Discussion}
General results, the method to unveil the most massive central cluster SMBHs and the outlook are part of this section.

\subsection{How to unveil the most massive black holes and implications for scaling relations}

In \S~\ref{Models} we compiled a set of SMBH growth rates which are based on the mechanical feedback energies of
central cluster AGN. The adiabatic growth of hot cluster gas results in a cusp regrowth process if the ratio of mass gain
to initial SMBH mass (at $z=1$) exceeds 25\%. This ratio will be lowered if the initial slope is smaller than $\gamma=0.2$
or more generally if the total mass within the core is smaller than assumed for our computations. Nevertheless, SMBH
growth in ordinary elliptical galaxies/BCGs (represented by our Models~1-4) should not lead to a
significant adiabatic contraction. This may change in the most massive and X-ray luminous clusters with the highest
feedback energies. Their central cluster black holes likely grow by several billion solar masses since redshift one unless
most of the binding energy is radiated away ($\epsilon\approx 1$). Assuming accretion efficiencies even below
$\epsilon\approx 0.1$, sources with the highest feedback energies might even double their masses during the hot cluster
gas accretion phase. Therefore, some nuclear black holes in the hottest and most X-ray luminous
clusters might have accumulated (through hot gas accretion and SMBH coalescences) masses of the order of
$\unit{10^{11}}{\msun}$ (see Section~\ref{Candidates} for some candidates). If the observation based models
(Equation~\ref{fmod_high}) with small $\epsilon$ are realized in some of these galaxy clusters, even the most prominent
cores will be healed or even replaced by steep density cusps. In Model~8 the initial core-S\'{e}rsic
model is transformed back into a shape which resembles a classical S\'{e}rsic profile whereas Model~9 develops a density cusp. 
This has consequences for the prediction power of the locally-calibrated $\mbh-R_{\mbox{\scriptsize{break}}}$,
$\mbh-R_{\mbox{\scriptsize{cusp}}}$ and $\mbh-L_{\mbox{\scriptsize{def}}}$ scaling relations. Black holes at the upper mass  limit  
do not necessarily inhabit galaxies with the most pronounced light deficits/cores. It could also affects the
$M_{V}-R_{\mbox{\scriptsize{cusp}}}$ relation \citep{2007ApJ...662..808L} which links the absolute V-band magnitude of a
galaxy to the size of the cusp radius obtained by a Nuker-profile fit. As light deficits are decreased by adiabatic contraction the
most massive black holes might be located in BCGs with cored-density profiles which are less pronounced than expected from
the total luminosity of their host galaxy. First observational support for our prediction might already have been found in
\cite{2015MNRAS.446.2330S}. However, it is necessary to include additional, even more extreme cases to see whether their
observed trend continues.

Finally, one can also argue that BCGs with huge break radii and light deficits like MS0735 which are know to be in
a rapid mass accretion process (unless $\epsilon\approx 1$) \citep{2009ApJ...698..594M} must indeed be ultramassive,
otherwise the ratio of $\Delta\mbh$ to the initial black hole mass would be too large and the core would have been replaced by a much
steeper cusp. In this sense our computations also strengthen the case that galaxies with huge central cores witness
the presence of ultramassive black holes. The ratio of black hole growth by merging (carving out shallow cores) to
hot gas accretion (steepening core profiles) is crucial.

\subsection{Massive black hole candidates and a test}\label{Candidates}

Three promising galaxy clusters which might host very massive black holes are RX J1347.5-1145,
SPT-CLJ2344-4243 (the Phoenix cluster) and Abell~2029. In this paragraph we will discuss how they are related to
our results and why they should be used for constraining hot gas accretion models. Additionally, they might help
to uncover the limitations of various galaxy-SMBH scaling relation. 

\begin{itemize}
\item \textbf{RX J1347.5-1145:} This object at redshift $z=0.45$ is one of the most
X-ray luminous ($L_{\mbox{\scriptsize{X}}}=\unit{6\cdot10^{45}}{\mbox{~erg~s}^{-1}}$ in the
[2-10] keV band) and massive clusters known \citep{2004A&A...427L...9G}. Depending on the accretion
efficiency and assuming that its mechanical AGN feedback balances central gas cooling as supported by
\citep{2002MNRAS.332..729C, 2007A&A...472..383G}, its nuclear black hole might have swallowed
several $\unit{10^{9}}{\msun}$ of hot gas (or even more if $\epsilon<0.1$, \S~\ref{growth_rate_ob}).
However, this is only one aspect. In their lensing analysis \citep{2014MNRAS.437.1858K} used 3D
pseudo-Jaffe models to parameterize the two central galaxies. For the brightest cluster galaxy they
deduced a velocity dispersion around $\sigma\approx\unit{600}{\mbox{km~s}^{-1}}$ (a value which
is also supported by gas spectroscopy but which might not necessarily result from a virialized state
\citealt{1998ApJ...492L.125S}) and a 3D core radius in excess of a few kpc (it will be reduced
in a 2D projection). If future observations confirm these values the case for an extremely massive
black hole in the core of RX J1347.5-1145 would be strengthened. It would require a black hole merger
rate at the upper end of the predictions of cosmological rich-cluster simulations \citep{2015MNRAS.451.1177L}
to carve out a kpc sized core in a galaxy with a huge velocity dispersion\footnote{The
black hole radius of influence, $R_{\mbox{\scriptsize{infl}}}$, decreases with growing velocity dispersion
since $R_{\mbox{\scriptsize{infl}}}\propto\sigma^{-2}$. Also note, that the
simulated clusters in \cite{2015MNRAS.451.1177L} are several times less massive (even) at $z=0$ than
RX J1347.5-1145 and SPT-CLJ2344-4243 at $z\approx0.5$.} and to
protect it against core healing by hot gas accretion.
\item \textbf{Phoenix cluster:} This cluster is even more massive and X-ray luminous than
RX J1347.5-1145 \citep{2012Natur.488..349M}. It could be one of the few objects in the visible universe
where a model as extreme as our Model~9 might be realized and whose nuclear black hole might exceed
$\unit{10^{11}}{\msun}$. Based on this scenario and our initial $N$-Body setup we predict that the old/red
stellar density profile of the Phoenix BCG (red curves in Figure~\ref{Dens_muesli}) should resemble a classic
S\'{e}rsic $n=4$ model which becomes slightly steeper and more cusp-like within $0.13R_{e}$. Interestingly, 
the inner most data points (within $\approx0.1R_{e}$) of the measured surface brightness profile in
Figure~3 from \cite{2013ApJ...765L..37M} seem to lie above the S\'{e}rsic $n=4$ reference model. 
However, note that our $N$-Body models are not fine-tuned to exactly match the properties of the Phoenix
BCG nor its redshift. A more compact size of its BCG of $R_{e}=\unit{17}{\mbox{ kpc}}$ \citep{2013ApJ...765L..37M}
but otherwise identical initial parameters results in a larger initial mass/light deficit. This would necessitate
even larger amounts of swallowed mass, $\Delta\mbh$, in order to steepen its central slope.
Likewise, the applied bulge mass of its BGC has an influence too, as shown by the two red curves
in Figure~\ref{Dens_muesli}.
\item \textbf{Abell~2029:} The central host galaxy, IC1101, is listed as one of the
optically most luminous and largest cD-galaxy \citep{1979ApJ...231..659D, 2004ApJ...617..879L}.
Therefore, a merger driven growth of its central black hole at the upper limits ($\mbh\approx\unit{\left[5-
7\right]\cdot10^{10}}{\msun}$) of cosmological predictions \citep{2015MNRAS.451.1177L} seems reasonable.
Additionally, IC1101 is embedded within a massive and relaxed galaxy cluster \citep{2003ApJ...586..135L}.
If classical Bondi accretion with its quadratical SMBH mass dependency (\citealt{1952MNRAS.112..195B},
our Section~\ref{A_Models}) is realized, the nuclear SMBH should accrete at high levels.
A precisely measured core size/light deficit might help to measure the ratio of SMBH
scouring to adiabatic growth in one of the most extreme galaxies.
\end{itemize}

Finally, \cite{2014ApJ...795L..31L} report on the BCG in the Abell~85 cluster. So far it is the
most extreme representative of a core-type galaxy with a measured break radius (cusp radius) of $R_{b}=
\unit{20}{\mbox{ kpc}}$ ($R_{c}=\unit{4.6}{\mbox{ kpc}}$) \citep{2014ApJ...795L..31L}\footnote{
Note that \cite{2015ApJ...807..136B} take a different position by showing that its stellar profile can be well
fitted with a classical S\'{e}rsic model without the need for a depleted core.} Based on the extrapolation of
the $\mbh-R_{\mbox{\scriptsize{break}}}$, $\mbh-R_{\mbox{\scriptsize{cusp}}}$ and $\mbh-
L_{\mbox{\scriptsize{def}}}$ relations, its central black hole might also have a mass around $\unit{10^{11}}{\msun}$
\citep{2014ApJ...795L..31L}.

It would be highly informative and important for constraining the limitations of gas accretion- and
SMBH merging processes, especially with respect to their cluster environments, if the nuclear black hole masses in
these four objects could be measured and compared with empirical scaling relations. In this way our prediction
that the truly most massive black holes are not necessarily located in galaxies with the largest cores could be put
on its first test.

\subsection{Outlook}\label{CAO}
We now discuss future applications with the black hole mass gain sensitive 'calorimeter', how its sensitivity
can be increased and describe a non-feedback regulated SMBH growth channel
which under right circumstances can be realized in the hottest and most massive galaxy clusters.

\subsubsection{Future applications}
A robust calorimeter requires detailed information about initial conditions. With a statistically meaningful
knowledge about central surface brightness profiles of BCGs at the onset of hot gas accretion, in combination
with deposited AGN feedback energies, the calorimeter can be used to determine the significance of cold- (during the quasar phase)
and hot gas accretion as well as SMBH merging. It also offers a tool to measure the accretion efficiency parameter.
The calorimeter method is based on photometric- instead of spectral-imaging, therefore it is much easier to apply than
the direct measurement of SMBH masses at these redshifts. A precisely calibrated SMBH mass gain sensitive
calorimeter offers the potential to constrain the growth history of individual galaxies and their central black holes.

We want to illustrate this idea on the basis of NGC~4889 and NGC~4874, the central dominating galaxies in the Coma
cluster. While NGC~4889 is indeed the brightest member, it has a nearly flat density core with a slope $\gamma=0.03$ and a break
radius of $R_{b}=\unit{970}{\mbox{ pc}}$ \citep{2007ApJ...664..226L}. Due to the flatness of its innermost surface
profile, we do not expect, on the basis of the results presented in this paper, a significant phase of hot gas accretion
throughout its life. Its central black hole of $\mbh=\unit{\left(6-37\right)\cdot10^{9}}{\msun}$ 
\citep{2011Natur.480..215M} was likely grown during a massive quasar phase whereas subsequent
SMBH merging activity has later shaped its core. NGC~4874 on the other hand has a much more diffuse appearance and
is thus classified as a cD-type galaxy. If we apply our results to this galaxy, we find a different history. Its central black
hole was not formed during a massive quasar phase but instead by the merger of several very massive progenitor galaxies.
This gave rise to its extended halo as well as a very large break radius of $R_{b}=\unit{1730}{\mbox{ pc}}$
\citep{2007ApJ...664..226L} formed by SMBH scouring. Its central cusp profile, $\gamma=0.12$, is steeper than that of
NGC~4889. Assuming its initial profile was flatter in the past (e.g. comparable to that of NGC~4889), hot gas accretion
was likely relevant in this galaxy. The recent infall of the galaxy group dominated by NGC~4889 \citep{2007A&A...468..815G}
maybe quenched its activity through the disruption of the accretion flow. We can estimate the central SMBH mass in
NGC~4874 from its break radius, the empirical $\mbh-R_{b}$ relation given in \cite{2013AJ....146..160R} and add
the mass required to increase its central slope by $\Delta\gamma=0-0.09$ (taken from our Model~7). However, the
value $\Delta\mbh[M_{9}]\approx10$ relevant for the highest increase $\Delta\gamma=0.09$ must been taken as an
upper limit: The measured slope in NGC~4874 is even flatter than $\gamma=0.2$ (which is chosen for our galaxy
computations) and the dynamical mass within its core is smaller (due to its more extended size and flatter slope) than assumed
for our reference Model~7. Consequently, even moderate hot gas accretion with a value around $\Delta\mbh[M_{9}]\approx1$ might
have shaped the central slope of NGC~4874. Its nuclear black hole might therefore be as massive as
$\mbh\approx\unit{\left(20-30\right)\cdot10^{9}}{\msun}$ with a tendency towards the lower value, making it
comparable to its neighbor. This mass estimate is also compatible with dynamical mass measurements performed
in \cite{1998AJ....115.2285M}. 

\subsubsection{Potential oscillations}
In our computations we assume that initial core formation proceeds via well established SMBH merger scenarios and that the
cores are fully grown at the onset of hot cluster gas accretion. Contrary to this assumption \cite{2012MNRAS.422.3081M,
2013MNRAS.432.1947M} show, that based on idealized computations (e.g. final SMBH masses following $\mbh-\sigma$),
centrally flattened profiles can also be produced by oscillations of the central potential induced by
repetitive AGN outbursts. However, their models predict core sizes much in excess of those typically observed
\citep{2014MNRAS.444.2700D}. Nevertheless, we hope that future studies which aim to investigate the influence of hot cluster
gas accretion onto the most massive black holes in giant cool core galaxy clusters consider both scenarios:
(i) Adiabatic cusp formation by hot gas accretion which is not restricted by the $\mbh-\sigma$ relation (see e.g.
\cite{2012MNRAS.424..224H} and the results obtained in this paper) and (ii) the destruction of density cusps
by potential oscillations. Without the cold clumps which are used in \cite{2013MNRAS.432.1947M} to mimic infalling
galaxies and whose extended gas halos might be stripped more efficiently in hotter and more massive galaxies clusters,
core sizes might be reduced due to less violent potential fluctuations \citep{2013MNRAS.432.1947M}. With smaller
initial cores created by potential oscillations, adiabatic cusp formation by hot gas accretion might dominate at least
in the most massive clusters hosting the most violently growing SMBHs.
Core creation by potential fluctuations can also be implemented in future versions of the \textsc{Muesli} software. 
In addition to the stellar/dark matter (DM) profile a particle distribution which represents the density profile of the ambient gas can
be generated in equilibrium. These gas particles are not dynamically evolved forward in time but their masses are
temporarily lowered within (e.g. bipolar) volumes to mimic cavities whose size depend on the equation of state of the gas and the jet power. 
Depending on additional parameters like the duty cycle of the AGN, longevity of cavities and the fraction of jet energy which goes
into cavity production and shock generation, the stellar/DM particles (which are dynamically evolved forward in time) will feel these
potential oscillation and react correspondingly. In addition to that the mass of the black hole is adiabatically increased as
already studied in this paper. The implemented angular base functions (spherical harmonics) already have a topology which
strongly resembles observed cavity systems. Furthermore, the high resolution scale of \textsc{Muesli} allows
to study the impact of potential oscillations deeply within the influence radius of the black hole, making it a reliable alternative to
grid based codes.

\subsubsection{SMBH runaway growth}
In Section~\ref{A_Models} we compared observed AGN feedback energies with those ones obtained from
the prediction of classical Bondi models. For the sake of simplicity we assumed that the growing black hole has no influence
on the accretion rate over typical cluster lifetimes of 8 Gyr. However, due to increasing evidence for the existence of black
holes in excess of $\unit{10^{10}}{\msun}$ and because of the generic $\mbox{d}\mbh/\mbox{d}t=\gamma\mbh^{2}$
dependency of classical Bondi accretion, this assumption might turn out to be wrong -- at least for a few extreme objects.
From a formal perspective, Bondi accretion should amplify itself and accelerate as:
\begin{equation}\label{Bondi_runaway_formula}
\mbh\left(t \right)=\frac{\mbh\left(t=0 \right)}{1-\gamma \mbh\left(t=0 \right)t}.
\end{equation}
However, care has to be taken. First of all (i), the gas reservoir is finite, (ii) the Bondi accretion parameters which
are expressed as $\gamma$ are not necessarily constant over time subject to AGN feedback and (iii) by deriving the
classical Bondi accretion rate (Equation~\ref{formula_bondi_classic}) one has to assume that the mass within the loss cone
is small compared to the mass of the accretor. Nevertheless, by using Equation~\ref{Bondi_runaway_formula} and the initial
parameters from Section~\ref{A_Models} with a black hole mass around $\unit{4\cdot10^{10}}{\msun}$, the accretion rate
is expected to accelerate. After several billion years it would exceed a few percent of the Eddington rate where it likely
changes to a radiatively efficient accretion state \citep{2003MNRAS.344...60G, 2003A&A...409..697M, 2004A&A...414..895F,
2008ApJ...687..156W}. Depending on the broad-band spectrum of the quasar\footnote{Thanks to
our referee for pointing this out.} and the gas temperature of the cluster it might either lead to inverse Compton cooling
\citep{1990MNRAS.247..439F} which decreases the gas temperature and would stimulate even higher accretion rates or
to Compton heating \citep{2004MNRAS.347..144S, 2005MNRAS.358..168S}. Interestingly, powerful quasars with luminousities
around $L\approx\unit{10^{47}-10^{48}}{\mbox{~erg~s}^{-1}}$ (comparable to our model predictions) are reported to be
located in the Phoenix \citep{2013ApJ...778...33U} and CL1821+643 cluster \citep{2014MNRAS.442.2809W}.
These are two of the strongest cool-core and most massive galaxy clusters.

\section{Summary}\label{Summary}
This paper is about a new strategy to unveil the most massive black holes in the universe. SMBHs at the highest mass scale
are important for constraining the limitations of empirical scaling relations which are useful for theoretical model building.
With growing evidence that the most massive black holes are located at the centers of very massive galaxy
clusters (e.g. \citealt{2012MNRAS.424..224H}) we present a simple and effective method to pick out the most extreme candidates. 

For that purpose we first transformed (observed) AGN feedback powers into central cluster SMBH growth rates
(Section~\ref{Models}). Through a comparison with classical Bondi models we checked whether the initial black hole masses adopted
for our $N$-Body computations are reasonable. We found a strong overlap between theoretically predicted and observationally
observed feedback powers. This indicates that simple Bondi accretion (even when adapted for rotating gas atmospheres) can
fuel the most powerful cluster AGN.

In the main part of our paper (Section~\ref{Computations_Results}) we computed with the help of the \textsc{Muesli}
software the dynamical response of the host galaxy to the growing SMBH. In this way we
constructed a black hole mass gain sensitive calorimeter. In principle, a precisely calibrated ``calorimeter''
offers a way to measure the contribution of cold gas accretion, SMBH merging and subsequent hot cluster gas
accretion to the growth of the central cluster black hole. For our initial $N$-Body setup we used core-S\'{e}rsic models 
to account for SMBH scouring as well as nine different SMBH growth scenarios, encompassing in logarithmically increasing
steps the whole parameters space of compiled gas accretion models.
We found that adiabatically driven cusp formation is significant in the most extreme growth models, but can be neglected if
the ratio of accreted hot gas to initial black hole mass is around $25\%$ or smaller. In this way our results also confirm the
expectation that galaxies with large central core-profiles (especially if they are located in galaxy clusters with huge deposited
feedback energies and by assuming they are not mainly spin powered) witness the presence of extremely
massive black holes. Otherwise the core would have been replaced by a steeper central slope.
However, in the most X-ray luminous galaxy clusters with permanent AGN feedback powers as high as 
($P_{\mbox{\scriptsize{AGN}}}\approx\unit{10^{46}}{\mbox{~erg~s}^{-1}}$) even the most pronounced initial cores
will be turned into cusps if accretion efficiencies are smaller than $\epsilon=0.1$. We argued that the nuclear black holes
inside RX J1347.5-1145, the Phoenix cluster and Abell~2029 might be among the most massive ones in the universe.
Secure mass measurements of their central black holes would not only help to uncover limitations of scaling relation but could
be used to constrain the limitations of gas accretion- and SMBH merging processes. They also might represent promising
candidates for future applications with the Event Horizon Telescope \citep{2009astro2010S..68D}. The hunt for Gargantua is on.

\begin{acknowledgements}
We thank Walter Alef and Helge Rottmann for their technical support with the VLBI-computer cluster and Olaf Wucknitz and
Vladimir Karas for helpful comments. Special thanks go to Stefanie Komossa for her readiness to improve the quality
of this manuscript and for the instructive discussions about black hole physics.
\end{acknowledgements}

\bibliographystyle{aa}

\end{document}